\documentstyle[12pt,epsf]{article}

\title{5D gravity and the discrepant G measurements}

\author{J.  P.  Mbelek \\ Service d'Astrophysique, C.E.  Saclay \\ F-91191
Gif-sur-Yvette Cedex, France}

\begin{document} \date{} \maketitle \baselineskip=8mm

\begin{abstract} It is shown that 5D Kaluza-Klein theory stabilized by an
external bulk scalar field may solve the discrepant laboratory G measurements.
This is achieved by an effective coupling between gravitation and the
geomagnetic field.  Experimental considerations are also addressed.
\end{abstract}

\section{Introduction}

Although the methods and techniques have been greatly improved since the late
nineteenth century, the precision on the measurement of the gravitational
constant, G, is still the less accurate in comparison with the other fundamental
constants of nature \cite{Luo}.  Moreover, given the relative uncertainties of
most of the individual experiments (reaching about $10^{-4}$ for the most
precise measurements), they show an incompatibility which leads to an overall
precision of only about 1 part in $10^{3}$.  Thus the current status of the $G$
terrestrial measurements (see \cite{Gillies}) implies either an unknown source
of errors (not taken into account in the published uncertainties), or some new
physics \cite{Melnikov}.  In the latter spirit, many theories which include
extradimensions have been proposed as candidates for the unification of physics.
As such, they involve a coupling between gravitation and electromagnetism (GE
coupling), as well as with other gauge fields present.\\

Here we show that the discrepancy between the present results of the
G-measurements may be understood as a consequence of the GE coupling.  Also,
this theory predicts a variation of the effective fine structure "constant"
$\alpha$ with the gravitational field, and thus with the cosmological time (see
\cite{paper 1}).

\section{Theoretical background}

An argument initially from Landau and Lifshitz \cite{Landau} may be applied to
the pure Kaluza-Klein (KK) action (see \cite{Mbelek}):  the negative sign of the
kinetic term of the five dimensional (5D) KK internal scalar field, $\Phi$,
leads to inescapable instability.  The question to know which of the two
conformally related frames (Einstein-Pauli frame or Jordan-Fierz frame) is the
best remains debated in the literature, each frame having its own advantage.
The negative kinetic energy density for the $\Phi$-field, and thus the
instability, occur in both frames.  In the following, we perform calculations in
the Jordan-Fierz frame, where, as we show, the discrepant laboratory
measurements of G find a natural explanation.  Also, it is true that the 5D KK
theory may yield a zero kinetic term (and thus a zero kinetic energy density,
which would also be unusual in 4D), but this occurs only when the
electromagnetic (EM) field is identically zero everywhere.  This may be relevant
for some cosmological solutions, but not for our discussion.  On the contrary,
we argue here that the $\Phi$-field is tightly related to the EM field via the
link between the compactified space of the fifth dimension and the U(1) gauge
group.

Since stabilization may be obtained if an external field is present, we assume
here a version, KK$\psi$, which includes an external bulk scalar field minimally
coupled to gravity (like the radion in the brane world scenario).  After
dimensional reduction ($\alpha = 0, 1, 2, 3$), this bulk field reduces to a four
dimensional scalar field $\psi = \psi (x^{\alpha})$ and, in the Jordan-Fierz
frame, the low energy effective action takes the form (up to a total divergence)
$$ S = - \,\int \sqrt{-g} \,\, [ \,\frac{c^{4}}{16\pi} \,\frac{\Phi}{G} \,R
\,\,+ \,\,\frac{1}{4} \,\,{\varepsilon}_{0} \,{\Phi}^{3} \,F_{\alpha\beta}
\,F^{\alpha\beta} \,\,+ \,\,\frac{c^{4}}{4\pi G} \,\frac{{\partial}_{\alpha}
\Phi \,{\partial}^{\alpha} \Phi}{\Phi} \,] \,d^{4}x $$ \begin{equation}
\label{KK action and Psi J-F frame} \,+ \,\int \sqrt{-g} \,\,\Phi \,[
\,\frac{1}{2} \,{\partial}_{\alpha} \psi \,\,{\partial}^{\alpha} \psi \,\,-
\,\,U \,\,- \,\,J \psi \,] \,d^{4}x, \end{equation} where $A^{\alpha}$ is the
potential 4-vector of the EM field, $F_{\alpha\beta} = {\partial}_{\alpha}
\,A_{\beta} \,- \,{\partial}_{\beta} \,A_{\alpha}$ the EM field strength tensor,
$U$ the self-interaction potential of $\psi$ of the symmetry breaking type and
$J$ its source term.

The source term of the $\psi$-field, $J$, includes the contributions of the
ordinary matter, of the EM field and of the internal scalar field $\Phi$.  For
each, the coupling is defined by a function (temperature dependent, as for the
potential $U$) $f_{X} = f_{X}(\psi\,,\,\Phi)$, where the subscript $X$ stands
for "matter", "EM" and "$\Phi$".  In addition, the necessity to recover the
Einstein-Maxwell equations in the weak fields limit, implies the following
conditions:  $U(v) = 0$ and $f_{EM}(v\,,\,1) = f_{matter}(v\,,\,1) = 0$, where
$v$ denotes the vacuum expectation value (VEV) of the $\psi$-field.

The contributions of matter and $\Phi$ are proportional to the traces of their
respective energy-momentum tensors.  A contribution of the form
${\varepsilon}_{0}\,f_{EM} \,F_{\alpha\beta} \,F^{\alpha\beta}$ accounts for the
coupling with the EM field.  Besides, though the fit to the data involves
$\frac{\partial f_{EM}}{\partial\Phi}\,(v\,,\,1)\,v \gg 4\pi G/c^{4}$, we may
infer that $\frac{\partial f_{EM}}{\partial\Phi}\,(v\,,\,1)\,v$ is negligibly
small at very high temperature ({\sl e.g.}, like in the core of the Sun or at
the big bang nucleosynthesis) and even vanishes beyond the critical temperature
(say $T_{c} = 6000$ K) of the potential $U = U(\psi, T)$ as one gets $v = 0$ in
that case.

Following Lichnerowicz \cite{Lichnerowicz}, let us interpret the quantity
\begin{equation} \label{G eff} G_{eff} = \frac{G}{\Phi} \end{equation} of the
Einstein-Hilbert term, and the factor ${\varepsilon}_{0eff} = {\varepsilon}_{0}
\,[ \,{\Phi}^{3} \,+ \,4 f_{EM}(\psi\,,\,\Phi) \,]$ of the Maxwell term
respectively as the effective gravitational "constant" and the effective vacuum
dielectric permittivity.  The effective vacuum magnetic permeability reads
${\mu}_{0eff} = {\mu}_{0} \,[ \,{\Phi}^{3} \,+ \,4 f_{EM}(\psi\,,\,\Phi)
\,]^{-1}$, so that the velocity of light in vacuum remains a true universal
constant.  Both terms depend on the local (for local physics) or global (at
cosmological scale) value of the KK scalar field $\Phi$, assumed to be positive
defined.

The least action principle applied to the action (\ref{KK action and Psi J-F
frame}) yields the generalized Einstein-Maxwell equations and the scalar fields
equations \begin{equation} \label{ext.  scalar field eq} {\nabla}_{\nu}
{\nabla}^{\nu} {\psi} = - \,\,J \,\,- \,\,\frac{\partial J} {\partial {\psi}}
\,\psi \,\,- \,\frac{\partial U} {\partial {\psi}} \end{equation} and
\begin{equation} \label{KK scalar eq} {\nabla}_{\nu}\,{\nabla}^{\nu}\,\Phi = -
\,\frac{4\pi G}{c^{4}} \,{\varepsilon}_{0}
\,F_{\alpha\beta}\,F^{\alpha\beta}\,{\Phi}^{3} \,\,+ \,\,U\,\Phi \,\,+ \,\,J
\psi\,\Phi \,\,+ \,\,\frac{\partial J} {\partial {\Phi}} \,{\Phi}^{2} \,\psi
\,\,- \,\,\frac{1}{2} \,( \,{\partial}_{\alpha} \psi \,\,{\partial}^{\alpha}
\psi \,) \,\Phi.  \end{equation}

\section{Solutions in presence of a dipolar magnetic field}

Let us study the {\sl spatial} variation of $\Phi$ in the weak fields conditions
out of the fields' source, but in presence of a static dipolar magnetic field,
$\vec{B} = \vec{\nabla}\,V(r, \varphi, \theta)$.  We denote $r$, $\varphi$ and
$\theta$ respectively the radius from the centre, the azimuth angle and the
colatitude.  Thus, writing $\Phi = \Phi (r, \varphi, \theta)$, and taking into
account that $\frac{\partial U}{\partial \psi} (v) = 0$ (definition of the VEV),
equation (\ref{KK scalar eq}) simplifies after linearization as \begin{equation}
\label{stabilized KK scalar eq stat.  lin.  approx} \Delta \Phi = -
\,\frac{2}{{\mu}_{0}} \,[ \,\frac{\partial f_{EM}}{\partial \Phi}\,(v\,,\,1)\,v
\,+ \,\frac{4\pi G}{c^{4}} \,] \,( \,\vec{\nabla} V \,)^{2}.  \end{equation}
Since $\Delta V = div\,\vec{B} = 0$ and $(\vec{\nabla} V)^{2} = \frac{1}{2}
\,\Delta (V^{2}) \,- \,V\,\Delta V$ identically, the solution of equation
(\ref{stabilized KK scalar eq stat.  lin.  approx}) above reads merely
\begin{equation} \label{stabilized KK scalar sol.  stat.  lin.  approx} \Phi = 1
\,- \,\frac{1}{{\mu}_{0}} \,[ \,\frac{\partial f_{EM}}{\partial
\Phi}\,(v\,,\,1)\,v \,+ \,\frac{4\pi G}{c^{4}} \,] \,V^{2}.  \end{equation}

For our purpose, it is sufficient to limit the expansion of the scalar
potential, $V$, to the terms of the Legendre function of degree one ($n = 1$)
and order one ($m = 1$).  Hence $V = (a^{3}/r^{2}) \,[ \,g^{0}_{1}\,\cos{\theta}
\,+ \,g^{1}_{1}\,\sin{\theta}\,\cos{\varphi} \,+
\,h^{1}_{1}\,\sin{\theta}\,\sin{\varphi} \,]$, where $g^{0}_{1}$, $g^{1}_{1}$
and $h^{1}_{1}$ are the relevant Gauss coefficients, $a$ is the Earth's radius
and $M = \frac{4\pi}{{\mu}_{0}} \,a^{3}\,\sqrt{(g^{0}_{1})^{2} \,+
\,(g^{1}_{1})^{2} \,+ \,(h^{1}_{1})^{2}}$ denotes its magnetic moment.  Setting
$\cos{{\varphi}_{1}} = -\,g^{1}_{1}/\sqrt{(g^{1}_{1})^{2} \,+
\,(h^{1}_{1})^{2}}$, $\sin{{\varphi}_{1}} = h^{1}_{1}/\sqrt{(g^{1}_{1})^{2} \,+
\,(h^{1}_{1})^{2}}$ and $\tan{\lambda} = g^{0}_{1}/\sqrt{(g^{1}_{1})^{2} +
(h^{1}_{1})^{2}}$, the solution of equation (\ref{stabilized KK scalar eq stat.
lin.  approx}) then reads (and similarly for $\psi$ by making the substitution
$\frac{\partial f_{EM}}{\partial \Phi} \rightarrow - \,\frac{\partial
f_{EM}}{\partial \psi}$) \begin{equation} \label{stabilized KK scalar sol.
stat.  1st approx} \Phi = 1 \,\,- \,\,\frac{1}{{\mu}_{0}} \,\frac{\partial
f_{EM}}{\partial \Phi}\,(v\,,\,1)\,v \,( \,\frac{{\mu}_{0}\,M}{4\pi\,r^{2}}
\,)^{2}\,x(\theta, \varphi), \end{equation} where we have set \begin{equation}
\label{def.  mixed variable x} x(\theta, \varphi) = \cos^{2}{\theta} \,\,+
\,\,\cot^{2}{\lambda}\,\sin^{2}{\theta}\,\cos^{2}{(\,\varphi \,\,+
\,\,{\varphi}_{1}\,)} \,\,- \,\,\cot{\lambda}\,\sin{2\theta}\,\cos{(\,\varphi
\,+ \,{\varphi}_{1}\,)}.  \end{equation} Thence, one derives the expression of
$G_{eff}(r, \theta, \varphi)$ by inserting the solution (\ref{stabilized KK
scalar sol.  stat.  1st approx}) above in relation (\ref{G eff}).

It is worth noticing that the magnetic potential, $V$, scales as $B \,r$, where
$B$ is the magnitude of the geomagnetic field at radius $r$.  Indeed, because of
this scaling effect, the small spatial variations of the geomagnetic field will
influence significantly the laboratory measurements of G whereas the large local
magnetic fields present in the laboratory ({\sl e.  g.}, the magnetic suspension
used to support the balance beam, the magnetic damper, etc...)  will not.  A
rough estimate shows that, even using a 30 Tesla superconducting magnet, one
still needs to gain at least one order of magnitude with the most precise G
measuring apparatus avalaible yet.  Hence, our prediction is consistent with the
earlier conclusion of Lloyd \cite{Lloyd}.

\section{Comparison with laboratory measurements} There are presently almost 45
results of G measurements published since 1942 \cite{paper 2}.  Because of the
too numerous uncontrolled systematic biases, the mine measurements are excluded
from the present study (including them will not change our fit because of their
lack of precision, typically less than $1\%$).  Also, the more discordant
laboratory measurement (high PTB value \cite{Braunschweig 95}) is excluded,
since it may suffer from a systematic effect.  The " official " values are
presently $G = 6.67259 \pm 0.00085~10^{-11}$ (CODATA 86, \cite{CODATA 86}) and
$G = 6.670 \pm 0.010~10^{-11}$ (CODATA 2000, \cite{CODATA 00}) in MKS unit.  In
the following, all the measurements are weighted equally in the fit.  Fitting
the 44 data with these values gives respectively $\chi^{2}_{\nu} = 11.128$ and
$\chi^{2}_{\nu} = 62.498$ ($\chi^{2}_{\nu} = \chi^{2}$ per degrees of freedom).
If we forget the official values and try a best fit, assuming an arbitrary
constant value of $G$, we obtain $G = 6.6741~10^{-11}~m^{3}~kg^{-1}~s^{-2}$ with
$\chi^{2}_{\nu} = 2.255$.  The fit to the same sample of 44 measurements (figure
1), on account of the GE coupling, yields (in MKS units) with $\chi^2_{\nu} =
1.669$ :  \begin{equation} \label{1/Glab vs x whole} \frac{1}{10^{11}~G_{eff}} =
( \,0.149929 \pm 0.000017 \,) \,\,- \,\,( \,0.0001509 \pm 0.0000252 \,) \,x(L,
l).  \end{equation} From the above fit, one derives both estimates of the true
gravitational constant \begin{equation} \label{G laboratory data} G = ( \,6.6696
\pm 0.0008 \,)~10^{-11}~m^{3}~kg^{-1}~s^{-2}, \end{equation} and the coupling
parameter \begin{equation} \label{coupling constant} \frac{\partial
f_{EM}}{\partial \Phi}\,(v\,,\,1)\,v = ( \,5.44 \pm 0.66 \,)~10^{-6}
\,fm\,\,TeV^{-1}.  \end{equation} The latter quantity, expressed in the
canonical form $\frac{\partial f_{EM}}{\partial \Phi}\,(v\,,\,1)\,v = \hbar c
\,M_{5}^{-2}$, yields a 5D Planck scale $M_{5} \simeq 5.9$~TeV of the order of
the value that is invoked in the literature to solve the hierarchy problem.

\begin{tabular}{|c|c|c|c|} \hline \emph{Location [reference]} & \emph{Latitude
($^{\circ}$)} & \emph{Longitude ($^{\circ}$)} & \emph{G$_{lab}$ \,($10^{-11}
\,m^{3} \,kg^{-1} \,s^{-2}$)} \\ \hline Lower Hutt (MSL) \cite{Lower Hutt 99,
Lower Hutt 95} & -41.2 & 174.9 & 6.6742 $\pm$ 0.0007 \\ & & & 6.6746 $\pm$
0.0010 \\ \hline Wuhan (HUST) \cite{Wuhan} & 30.6 & 106.88 & 6.6699 $\pm$ 0.0007
\\ \hline Los Alamos \cite{Los Alamos} & 35.88 & -106.38 & 6.6740 $\pm$ 0.0007
\\ \hline Gaithersburg (NBS) \cite{Gaithersburg 82, Gaithersburg 42} & 38.9 &
-77.02 & 6.6726 $\pm$ 0.0005 \\ & & & 6.6720 $\pm$ 0.0041 \\ \hline Boulder
(JILA) \cite{Boulder} & 40 & -105.27 & 6.6873 $\pm$ 0.0094 \\ \hline Gigerwald
lake \cite{Gigerwald lake 95, Gigerwald lake 94} & 46.917 & 9.4 & 6.669 $\pm$
0.005 (at 112 m) \\ & & & 6.678 $\pm$ 0.007 (at 88 m) \\ & & & 6.6700 $\pm$
0.0054 \\ \hline Zurich \cite{Zurich 98, Zurich 99} & 47.4 & 8.53 & 6.6754 $\pm$
0.0005 $\pm$ 0.0015 \\ & & & 6.6749 $\pm$ 0.0014 \\ \hline Budapest
\cite{Budapest} & 47.5 & 19.07 & 6.670 $\pm$ 0.008 \\ \hline Seattle
\cite{Seattle} & 47.63 & - 122.33 & 6.674215 $\pm$ 0.000092 \\ \hline Sevres
(BIPM) \cite{Sevres 01, Sevres 99} & 48.8 & 2.13 & 6.67559 $\pm$ 0.00027 \\ & &
& 6.683 $\pm$ 0.011 \\ \hline Fribourg \cite{Fribourg} & 46.8 & 7.15 & 6.6704
$\pm$ 0.0048 (Oct.  84) \\ & & & 6.6735 $\pm$ 0.0068 (Nov.  84) \\ & & & 6.6740
$\pm$ 0.0053 (Dec.  84) \\ & & & 6.6722 $\pm$ 0.0051 (Feb.  85) \\ \hline
Magny-les-Hameaux \cite{Magny-les-Hameaux} & 49 & 2 & 6.673 $\pm$ 0.003 \\
\hline Wuppertal \cite{Wuppertal} & 51.27 & 7.15 & 6.6735 $\pm$ 0.0011 $\pm$
0.0026 \\ \hline Braunschweig (PTB) \cite{Braunschweig 95, Braunschweig 87} &
52.28 & 10.53 & 6.71540 $\pm$ 0.00056 \\ & & & 6.667 $\pm$ 0.005 \\ \hline
Moscow \cite{Moscow 98, Moscow 79} & 55.1 & 38.85 & 6.6729 $\pm$ 0.0005 \\ & & &
6.6745 $\pm$ 0.0008 \\ \hline Dye 3, Greenland \cite{Dye 3} & 65.19 & -43.82 &
6.6726 $\pm$ 0.0027 \\ \hline Lake Brasimone \cite{lake Brasimone} & 43.75 &
11.58 & 6.688 $\pm$ 0.011 \\ \hline \end{tabular}\\\\Table 1 :  Results of the
most precise laboratory measurements of G published during the last sixty years
and location of the laboratories.\\

\begin{figure} \centerline{\epsfxsize=12cm \epsfbox{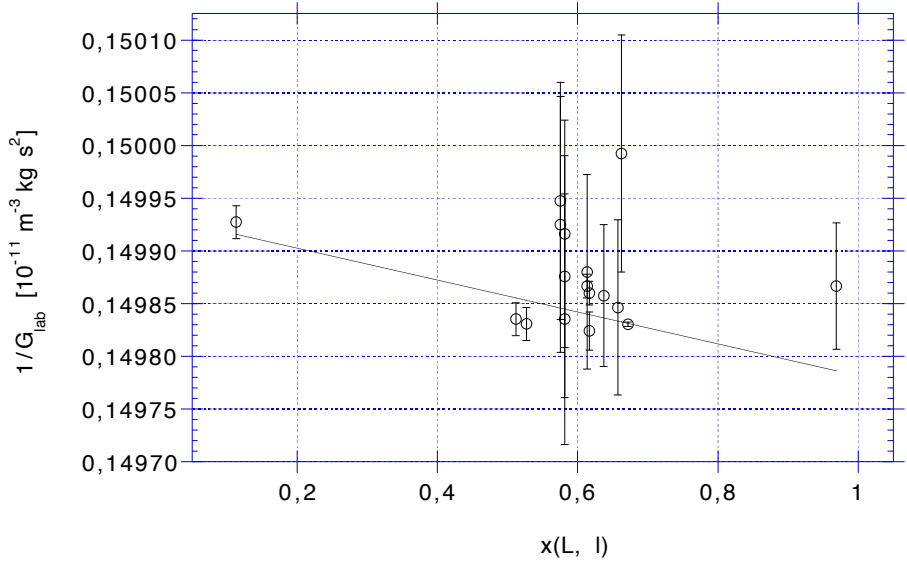}}
\caption{Laboratory measurements with relative uncertainty $\frac{\delta
G_{lab}}{G_{lab}} < 10^{-3}$ and measuring time $\Delta t < 200~s$ (sample S1,
17 points \cite{Lower Hutt 99}, \cite{Wuhan} - \cite{Gaithersburg 42},
\cite{Gigerwald lake 94}, \cite{Seattle}, \cite{Fribourg} - \cite{Wuppertal},
\cite{Braunschweig 87} - \cite{Dye 3}).  The line indicates the best fit
$G_{lab}$ versus the mixed variable $x$ ($\chi^{2}_{\nu} = 1.327$).  Assuming a
constant $G$ would yield a bad fit to the data ($\chi^{2}_{\nu} = 3.607$),
mostly because of the HUST value.}  \end{figure}

\begin{figure} \centerline{\epsfxsize=12cm \epsfbox{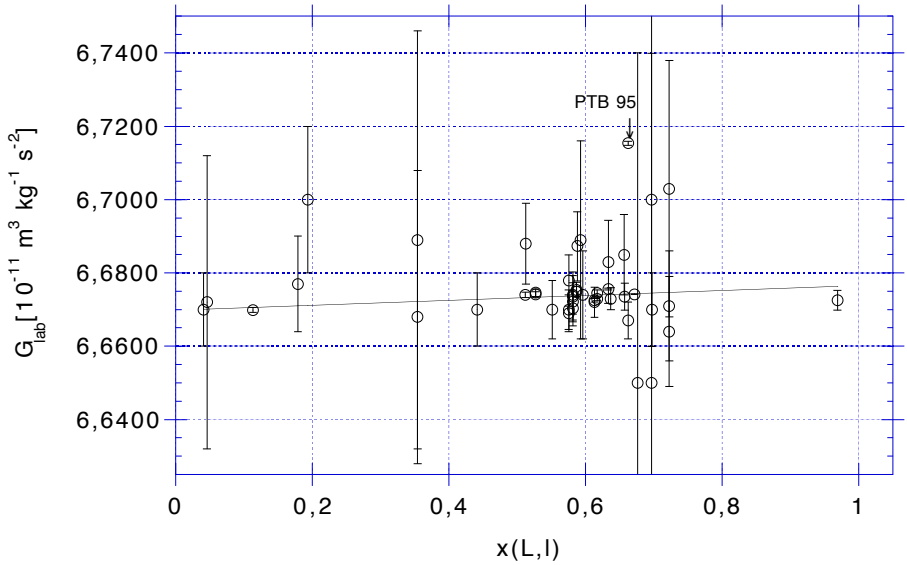}}
\caption{$G_{lab}$ versus $x$ (whole sample plus the PTB 95 value, 45 points
\cite{Lower Hutt 95} - \cite{others}).}  \end{figure}

\begin{tabular}{|c|c|c|} \hline Sample & H0 & H1 \\ \hline S1 & ~ &~ \\ 17
     points & $\chi^{2}_{\nu}$ = 3.607 (best fit) & \\ & $\chi^{2}_{\nu}$ =
     21.523 (mean of CODATA 86) & $\chi^{2}_{\nu}$ = 1.327\\ (Fig.1)&
     $\chi^{2}_{\nu}$ = 141.46 (mean of CODATA 2000) & \\ \hline Whole &
     $\chi^{2}_{\nu}$ = 2.255 (best fit) & \\ \cite{Lower Hutt 99} -
     \cite{others}& $\chi^{2}_{\nu}$ = 11.128 (mean of CODATA 86) & \\ 44 points
     & $\chi^{2}_{\nu}$ = 62.498 (mean of CODATA 2000) & $\chi^{2}_{\nu}$ =
     1.669 \\(Fig.2) &~ &~ \\ \hline \end{tabular}\\\\Table 2 :  Reduced
     $\chi^{2}$ for the two different hypothesis H0 (Hypothesis of a constant
     $G$) and H1 (Hypothesis of an effective G), and different samples S1 and
     whole (except the high PTB value \cite{Braunschweig 95}, see text).\\

Considering the whole sample, we check the relevance of our result under H1
compared to the best value of $G$ under H0, by applying the F test (Fisher law).
This yields $F_{\chi} = \frac{\Delta \chi^2}{\chi^2_{\nu}} = 16.09$, which
indicates that, independently of the number of parameters (two instead of one),
our fit is better with a significance level greater than $99.9\%$
\cite{Bevington}.  Let us emphasize that the most precise value of G today
\cite{Seattle} contributes to $\chi^{2}_{\nu}$ to less than $0.0006$ in the
above fit.  Likewise, the last published value of G \cite{Zurich} would
contribute to less than $0.02$.  This suggests that the agreement may be better
than purely indicated by the $ \chi ^{2}$ values.  Moreover, if one substitutes
$G = (6.6731 \pm 0.0002)~10^{-11}~m^{3}~kg^{-1}~s^{-2}$~\cite{Newman} announced
by the MSL team in June 2002 (CPEM, Ottawa, Canada) for the prior $G = (6.6742
\pm 0.0007)~10^{-11}~m^{3}~kg^{-1}~s^{-2}$ published in 1999~\cite{Lower Hutt
99}, one would obtain with the sample S1 :

H0 :  ${\chi}_{\nu}^{2} = 5.168$

H1 :  ${\chi}_{\nu}^{2} = 1.162$

and with the whole sample of 44 measurements :

H0 :  ${\chi}_{\nu}^{2} = 2.836$

H1 :  ${\chi}_{\nu}^{2} = 1.488$

\section{Discussion and Conclusion} It is worth noticing that the scalar fields
under considerations identify neither to the dilaton nor to the inflaton of
higher dimensional theories, without further assumptions.  In particular, the
computation of both scalar fields, as given by equations (\ref{ext.  scalar
field eq}) and (\ref{KK scalar eq}), involves only (see the right hand sides)
quantities related to fields sources, and not to the test bodies.  Hence, the
effective $G_{eff}$ given by relation (\ref{G eff}) does not depend on the
composition of the test bodies.  Besides, let us emphasize that the equation of
motion of a neutral point-like particle in the genuine KK theory reduces to the
4D geodesic equation after dimensional reduction~\cite{Wesson}, although the KK
scalar field is coupled to the Maxwell invariant $F_{\alpha\beta}
\,F^{\alpha\beta}$.  In a forthcoming paper~\cite{paper 3}, we address the
effect of the varying effective coupling constants on the masses of composit
particles.  On account of the Higgs mechanism of quarks and leptons masses
generation, by promoting the Yukawa coupling constants to effective parameters
(on an equal footing with $G$ or $\alpha$) that depend on both scalar fields
$\psi$ and $\Phi$, we prove that the KK$\psi$ model is actually consistent with
the current experimental bounds on the violation of the equivalence principle.
Hence, we conclude that present laboratory experiments may not measure a true
constant of gravitation.  Instead, in addition to all other possible biases
({\sl e.  g.}, anelasticity in the wire of torsion pendulum as pointed out by
Kuroda \cite{Kuroda}, and which since has been generically taken into account in
the experiments), they may be pointing out an effective one depending on the
geomagnetic field at the laboratory position.

\end{document}